\documentclass{article}

\newcommand{\product}[2]{ {#1} {\times} {#2} }

\newcommand{\pair}[2]{\langle #1,#2 \rangle}
\newcommand{\triple}[3]{\mbox{$ \langle #1,#2,#3 \rangle $}}

\title{\Large\bf Conceptual Analysis of Resource Meta-information}
\author{
  
  \small Christian Neuss \\
  \small neuss@igd.fhg.de \\
  \and
  \small Robert E. Kent \\
  \small rekent@logos.ualr.edu\\  
}
\date{}

\begin{document}
	\maketitle

\begin{abstract}

It's ease of use and the availability of browsers for various platforms
have paved the way for the enormous popularity 
that the World Wide Web currently enjoys.
In the near future,
by providing 
not only easy access to information,
but also means for conducting business transactions,
the Web could form the base technology for the information superhighway.
In such a large distributed information system,
resource discovery becomes a critical problem.

Recent developments in resource discovery systems,
such as {\sc Harvest} \cite{Harvest94b} and {\sc Whois}$+$$+$ \cite{DeWe94},
provide scalable mechanisms 
for the identification, location and characterization 
of networked information resources
based upon {\em resource meta-information\/}.
However,
the Web's vast information space can only be handled effectively,
when resources are meaningfully classified into coherent conceptual structures.

The automatic classification of resource meta-information
is at the heart of the {\sc wave} system \cite{WAVE94},
which employs 
methods from the mathematical theory of concept analysis 
to analyze and interactively explore 
the vast information space 
defined by wide area resource discovery services.
In this paper 
we discuss these methods
by interpreting various synoptic and summary interchange formats 
for resource meta-information,
such as the {\sc Harvest} {\tt SOIF} and the {\sc Whois}$+$$+$ {\tt urc},
in terms of basic ideas from concept analysis.
In so doing,
we advocate
concept analysis as a principled approach to effective resource discovery.

\end{abstract}

\section{Introduction}

The World Wide Web is an Internet based information system 
for access to distributed hyperlinked documents.
Its ease of use and the ability 
of seamlessly integrating other information services 
have paved the way for
the widespread popularity the Web currently enjoys.
Finding specific information however is a difficult task
that cannot be accomplished through browsing. 
To address this problem, 
various mechanisms for resource discovery have been developed.

A naming scheme,
such as the {\sc IETF} {\sc URN},
supports the concept of a persistent location-independent resource identifier. 
Directory resolution services 
map resource names ``back'' to actual physical locations
(resource instances).
By encapsulating resource meta-information together with resource names,
attributes,
such as ``title'', ``author'' or ``topic'', etc.\,
are available to search engines.
In order to implement sophisticated retrieval engines, 
a means for the automatic interpretation and classification of meta-information 
must be found. 
Furthermore, 
an adequate metaphor for
presenting and exploring information structures 
must be developed. 

This paper discusses 
recent developments in the area of information resource discovery services, 
and proposes the use of methods from the mathematical theory of concept analysis
to process and interactively browse a large information space.


\section{Resource Meta-information}

\subsection{Uniform Resource Characteristics}

We here want to apply ideas from Concept Analysis 
to the on-going discussions in the {\sc IETF-IIIR} and {\sc IETF-URI} working groups in general,
and the specification by Michael Mealling \cite{Me94a,Me94b} of {\sc URC}s in particular.
We use the following definitions.
\begin{itemize}
	\item A Uniform Resource Locator ({\sc URL}) 
		is used for hyperlink markup in Web documents. 
		Since 
		a {\sc URL} specifies a location and retrieval protocol of a given network resource,
		it is not a long-lived, stable reference.
		Moving or even renaming a file causes a {\sc URL} reference to become stale. 
		Besides, 
		copies of the same document can be located at multiple locations ({\sc URL}s). 
	\item A Uniform Resource Name ({\sc URN}) 
		is used to {\em identify\/} a resource.
		A Uniform Resource Name ({\sc URN}) 
		is an identifier 
		that uniquely and persistently names an information resource.
		The {\sc URN} scheme has been designed 
		in order to solve various problems with the {\sc URL}.
 		A Uniform Resource Locator ({\sc URL}) 
		is used to {\em locate\/} an instance of a resource
		identified by an {\sc URN}.
		A Uniform Resource Identifier ({\sc URI}) 
		is either a {\sc URN} or a {\sc URL}.
	\item A Uniform Resource Characteristic ({\sc URC}) 
		is used to {\em represent\/} {\sc URI}s 
		and their associated meta-information.
		Uniform Resource Characteristics ({\sc URC}s) \cite{Me94a}
		are analogous to the bibliographic records of Library Science.
		{\sc URC}s encode meta-information about network resources
		in the form of attribute$\colon$value pairs order by precedence.
		Compare also with
		{\sc IAFA} formats, {\sc Whois}$+$$+$ templates, 
		{\sc SOIF} of the {\sc Harvest} system for resource discovery, etc.
\end{itemize}

\subsection{IAFA Templates}

\subsection{Harvest Summary Object Interchage Format}

\subsection{Bibliographic Records from Library Science}

Below,
classified according to the eight areas of description of ISBD in Library Science,
we list some attributes which might be relevant for a particular purpose:
\begin{enumerate}
	\item title and statement of responsibility
		(Title, Author)
	\item edition
		(Version)
	\item material (or type of publication) specific details
	\item publication, distribution, etc.
	\item physical description
		(Content-Type, Content-Length, Size, Cost, etc.)
	\item series
		(Time To Live)
	\item notes
		(Abstract)
	\item standard number and terms of availability
		(Uniform Resource Names, Uniform Resource Locators)
\end{enumerate}



\section{Resource Discovery Services}

Due to the rapid growth of the World Wide Web in 1994, 
resource discovery has become a serious problem. 
Because of its decentralized architecture, 
the user experiences the Web as 
a large information repository without an underlying structure. 
The process of ``surfing'' pages by repeatedly following links 
is the most popular use of the Web.
It can however lead to the phenomenon of getting ``lost in hyperspace''.

\subsection{Walking the Web}

From the very beginning, 
approaches have been made 
to organize information about networked information resources 
into catalogs and indexes.
Index files were originally maintained manually.
However, 
the rapid growth of the Web soon made necessary
automatic methods for generating resource directories.
Automatic tools 
called ``robots'', ``Web wanderers'' or ``spiders'' 
soon evolved.
These are programs 
which automatically connect to a remote server 
and recursively retrieve documents. 
Since spider programs often put heavy loads on Web servers,
they have been controversial,
and are sometimes disliked by server maintainers.

The main problem with ``spiders'' is that they are nor ``true Web wanderers''
--- the retrieval program does not transfer itself 
from the index site to the provider site, 
but instead transfers 
in the reverse direction over the network
all the potentially indexable documents.
Since document repositories may contain hundreds of megabytes, 
the bandwidth requirements are enormous. 
Exacerbating this problem is the fact that
current indexing tools gather independently,
without sharing information with other indexers.


\subsection{Whois++-URC}

The {\sc WNILS}\footnote{
  {\sc Whois} and Network Information Lookup Service
} working group in the Internet Engineering Task Force ({\sc IETF}) 
is currently defining a standard 
for creation of a distributed directory service
called {\sc Whois}$+$$+$. 
It defines the notion of ``centroid''
as a mechanism for passing index information between 
index servers. 
A {\em centroid\/} is a list of 
records, attributes, and a word list for each attribute.
%
%
The word list for a given attribute 
contains one occurrence of every word 
which appears at least once in any record in the database for the attribute.
In order to optimize searching,
this abstracted information 
is passed up a hierarchical tree to other servers.
[QUOTE Gargano Proc.Inet'93]

Although {\sc Whois}$+$$+$ defines a general purpose directory service, 
it can be employed to provide a Web specific resource discovery service. 
This can be achieved through a {\sc URN} to {\sc URL} mapping directory.
By gathering and distributing {\sc URC} information 
through a hierarchy of {\sc Whois}$+$$+$ servers, 
resource directories based on attributes such as ``title'' and ``keywords'' 
become possible.

\subsection{Harvest}

{\sc Harvest} \cite{Harvest94a,Harvest94b}
is a distributed system for resource discovery and indexing. 
It separates the task of obtaining and distributing data: 
A {\it gatherer\/} collects information from a provider, 
while a {\it broker\/} provides a query interface for index requests.
This approach has a variety of advantages.
First of all, 
being able to run the gatherer at the provider site 
reduces server load and network traffic. 
Secondly,
since a gatherer can feed information to many brokers,
some of the redundency described in the previous section can be avoided. 
Not only can brokers access more than one gatherer,
but they can also be used to cascade indexed views from other brokers,
using their query interface.


Not unlike the {\sc Whois}$+$$+$ approach,
{\sc Harvest} uses a record consisting of attribute/value pairs 
as a unit for information indexing. 
The {\it Summary Object Interchange Format\/} ({\tt SOIF})
is based on a combination of 
IETF/IAFA 
templates
and 
the BibTeX format \cite{LaTeX86}. 
Figure~\ref{pSOIF} shows an example of an {\tt SOIF} template 
\cite{Harvest94b}.

\begin{figure}[htb]
\begin{footnotesize}
\begin{verbatim}
        @FILE {   ftp://sunsite.unc.edu/pub/packages/
          database/lincks/lincks-2.2.1.tar.gz
        Time-to-Live {7}:	9676800
        Last-Modification-Time {9}:	774648862
        Refresh-Rate {7}:	2419200
        ...
        Title {28}:	LINCKS - a multi-user OODBMS
        Version {17}:	2.2 patch level 1
        Description {383}:	LINCKS Sources and Documentation
        LINCKS is an object-centred multi-user database system
        developed for complex information system applications
        where editing and browsing of information in 
        the database is of paramount importance.        The focus
        is on sharing of small information chunks which combine
        to make up complex information         objects used 
        by different users for different purposes.
        Author {28}:	Lin Padgham, Ralph Ronnquist
        AuthorEmail {59}:	lincks@ida.liu.se, or linpa@ida.liu.se
        and ralro@ida.liu.se
        MaintEmail {17}:	lincks@ida.liu.se
        ...
\end{verbatim}
\end{footnotesize}
\caption{Excerpt from sample SOIF file}
\label{pSOIF}
\end{figure}


Table~\ref{analogies} lists some analogies 
between various components of resource discovery services.
It provides an orientation towards the specification of
resource discovery services
in a distributive fashion
as a federation of resource discovery software agents.

\begin{table}[htb]
\begin{center}
\begin{tabular}{|l|l|l|}\hline
	\multicolumn{1}{|c|}{\bf Library Science} 
		& \multicolumn{1}{|c|}{{\sc Whois}$+$$+$}
			& \multicolumn{1}{|c|}{\sc Harvest}
				 \\ \hline\hline
	{\bf patron}
		& user
			& user
				\\ \hline
	{\bf photo-copier}
		& {\tt HTTP} for resource
			& {\tt HTTP} for resource
				\\ \hline\hline
	{\bf reference librarian}
		& client/user-interface
			& Broker/Query-Manager
				\\ \hline
	{\bf card catalog}
		& index
			& Broker/Registry
				\\ \hline
	{\bf holdings manager}
		& Base-level Server
			& Broker/Storage-Manager
				\\ \hline
	{\bf interlibrary loan}
		& POLL command
			& Broker/Collector
				\\ \hline\hline
	{\bf cataloging}
		& individual publisher
			& Gatherer/Extractor(Essence)
				\\ \hline\hline
	{\bf circulation}
		& Directory Mesh (?)
			& Broker/Caching-and-Replicator
				\\ \hline\hline
	{\bf library resource list}
		& IANA
			& Harvest Server Registry (HSR)
				\\ \hline
\end{tabular}
\end{center}
\caption{{\bf Analogies between Resource Discovery Services}}
\label{analogies}
\end{table}


\section{Resources are Conceptual Classes}

Using ideas from Library Science and Concept Analysis,
we are currently developing tools for the conceptual analysis
of networked information resources in general,
and the World Wide Web in particular.
Networked information resources \cite{WeDe94} include
(1) individual text files,
(2) WAIS databases,
and
(3) starting points for hypertext webs.
In this section we argue by example
that resources are best thought of,
not as objects,
but as conceptual classes (concepts).
We offer a concept-oriented approach
for the description and organization of networked information resources,
which will facilitate their subsequent discovery and access.
This should not be thought of as yet another object-oriented approach.
Although objects generate their own classes,
classes are not only more general 
but also include intensional information.
By identifying concepts with classes,
this can be regarded as a class-oriented approach
--- an approach that has been advocated recently by Terry Winograd
in the IETF-URI working group discussion on library standards and {\sc URI},
and supported by Ronald Daniel and Dirk Herr-Hoyman.

\subsection{Concept Analysis}

Concept Analysis \cite{Wi82}
is a relatively new discipline
arising out of the mathematical theory of lattices and category theory.
It is closely related to the areas of 
knowledge representation in Computer Science 
and
Cognitive Psychology.
Concept Analysis provides for 
the automatic classification of both knowledge and documents
via 
representation of a users faculty for interpretation
as encoded in conceptual scales.
Such conceptual scales correspond to 
the facets of synthetic classification schemes,
such as Ranganathan's COLON scheme, in Library Science.

Concept Analysis uses objects, attributes and conceptual classes
as its basic constituents.
A {\em conceptual class\/} consists of any group of entities or {\em object\/}s
exhibiting one or more common characteristics, traits or {\em attribute\/}s.
A characteristic is a conceptualized attribute 
by which classes may be identified and separated into a conceptual hierarchy, 
and further subdivided (specialized) 
by the facets of topic, form, location, chronology, etc.
The ``has'' relationship between objects and attributes
is represented as a binary relation called a formal context.

A {\em formal context\/} is a triple $\triple{G}{M}{I}$
consisting of two sets $G$ and $M$
and a binary incidence relation $I \subseteq \product{G}{M}$ between $G$ and $M$.
Intuitively, 
the elements of $G$ are thought of as entities or objects,
the elements of $M$ are thought of as properties, characteristics or attributes
that the objects might have,
and $g{I}m$ asserts that ``object $g$ has attribute $m$.''
In many contexts appropriate for Web documents,
the objects are documents
and the attributes are any interesting properties of those documents.

The definition of a conceptual class must involve:
the common attributes,
which are encoded in the superordinate 
(next higher and more general class),
and 
the distinguishing attributes,
which differentiate the defined concept from the superordinate.
Conceptual classes are logically characterized by their extension and intension.
\begin{itemize}
	\item The {\em extension\/} of a class is 
		the aggregate of entities or objects
		which it includes or denotes.
	\item The {\em intension\/} of a class is 
		the sum of its unique characteristics, traits or attributes,
		which, taken together,
		imply the concept signified by the conceptual class.
\end{itemize}
The intent should contain precisely those attributes 
shared by all objects in the extent,
and vice-versa,
the extent should contain precisely those objects 
sharing all attributes in the intent.
Clearly the terms ``extension'' and ``intension'' are reciprocally dependent.
They complement each other by 
reciprocally deliminating concepts and explicating definitions.
A conceptual class will consist of such an extent/intent pair.

The process of subordination of conceptual classes and collocation of objects
exhibits a natural order,
proceeding top-down
from the more general classes with larger extension and smaller intension
to the more specialized classes with smaller extension and larger intension.
This order is called generalization-specialization.
One class is more specialized (and less general) than another class,
when its intent contains the other's intent,
or equivalently,
when the opposite ordering on extents occurs.
Conceptual classes with this generalization-specialization ordering
form a class hierarchy for the formal context.
Knowledge is here represented as the hierarchical structure,
or complete lattice,
known as the {\em lattice of conceptual classes\/} of the formal context.
Such lattices of classes can be visualized by line diagrams,
where nodes represent conceptual classes
and edges represent the subclass (subtype) relationship.

The join of a collection of conceptual classes
represents the common attributes or shared characteristics of the classes.
The bottom of the conceptual hierarchy (the empty join)
represents the most specific class 
whose intent consists of all attributes
and whose extent is often empty.
The meet of a collection of conceptual classes
represents the conjunction of all the attributes of the classes.
The top of the conceptual hierarchy (the empty meet)
represents the universal class whose extent consists of all objects.
The entire conceptual class hierarchy is implicitly specified
by the ``has'' relationship of the formal context.
However,
part of the hierarchy of conceptual classes 
could also be explicitly specified 
via the following top-down process.
\begin{itemize}
	\item {\bf Initialization:} 
		The main top-level attribute classes are specified.
		These are meet-irreducible classes,
		meaning that they cannot be expressed
		as the meet of other more general classes.
	\item {\bf Iteration:} 
		Any collection of (super)classes can be specialized
		by the specification of differentiating attributes,
		thus producing subclasses.
		Each such differentiated (sub)class 
		is subordinate to 
		every (super)class in the collection.
	\item {\bf Termination:}
		Continue until further specialization and differentiation
		is either impossible or impractical.
\end{itemize}


\subsection{Specification of Resource Meta-information}

Table~\ref{formats} lists two generic interchange formats 
which can be used to specify 
\underline{faceted} information in conceptually scaled networked information resources.
Such faceted information can occur in various interfaces in a resource discovery system.
From a mathematical viewpoint,
these two representations are equivalent to each other.
Software exists for converting between the two forms.
The left side of Table~\ref{formats} displays
the {\em Formal Context Interchange Format\/} ({\tt FCIF}).
{\tt FCIF} is oriented towards the formal contexts of Concept Analysis.
{\tt FCIF} represents
order-theoretic formal contexts of networked information resources,
consisting of
two partially ordered sets,
a poset of objects and a poset of single-valued attributes,
and an order-preserving incidence matrix
which represents the {\em has\/} relationship between objects and attributes.
The right side of Table~\ref{formats} displays
the {\em Concept Lattice Interchange Format\/} ({\tt CLIF}).
{\tt CLIF} is oriented towards the concept lattices of Concept Analysis.
{\tt CLIF} provides a storage-optimal representation of
order-theoretic lattices of conceptual classes for networked information resource meta-information,
consisting of (the inverse relationships for)
two generator monotonic functions,
from the posets of objects and attributes to the lattice of conceptual classes,
and a successor matrix
which represents the subtype relationship between conceptual classes.

\begin{figure}[htb]
\begin{center}
\begin{tabular}{cc}
\begin{minipage}{6cm}
\begin{center}
\begin{tabular}{|c|}\hline
{\bf Conceptual Scaling with} {\sc URC} \\ \hline\hline
\begin{picture}(150,40)(30,35)
		\put(48,63){{\tt FCIF}}
		\put(72,60){\vector(-1,0){30}}
		\put(48,43){{\tt CLIF}}
		\put(72,40){\vector(-1,0){30}}
	\put(72,50){\fbox{\begin{minipage}{60pt}
		\begin{tabular}{c}
			Conceptual \\
			Scaling
		\end{tabular}
	\end{minipage}}}
		\put(170,50){\vector(-1,0){30}}
		\put(150,53){{\tt URC}}
\end{picture} \\ \hline
\end{tabular}
\end{center}
\end{minipage}
&
\begin{minipage}{6cm}
\begin{center}
\begin{tabular}{|c|}\hline
{\bf Conceptual Scaling with} {\tt SOIF} \\ \hline\hline
\begin{picture}(150,40)(30,35)
		\put(48,63){{\tt FCIF}}
		\put(72,60){\vector(-1,0){30}}
		\put(48,43){{\tt CLIF}}
		\put(72,40){\vector(-1,0){30}}
	\put(72,50){\fbox{\begin{minipage}{60pt}
		\begin{tabular}{c}
			Conceptual \\
			Scaling
		\end{tabular}
	\end{minipage}}}
		\put(170,50){\vector(-1,0){30}}
		\put(146,53){{\tt SOIF}}
\end{picture} \\ \hline
\end{tabular}
\end{center}
\end{minipage}
\end{tabular}
\end{center}
\caption{{\bf Conceptual Scaling with Various Interchange Formats}}
\label{interchange}
\end{figure}
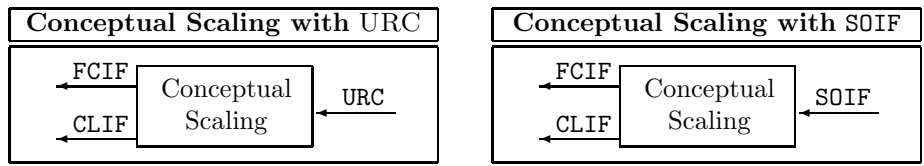

In this section we demonstrate that
{\tt FCIF} and {\tt CLIF} subsume
both the Uniform Resource Characteristics of {\sc Whois}$+$$+$
and the Summary Object Interchange Format of {\sc Harvest}.
These interchange formats are more general mechanisms than either {\sc URC}s or {\sc SOIF}s,
and allow for the specification of more complex conceptually structured systems of resources.
Actually,
as Figure~\ref{interchange} points out,
both {\tt FCIF} and {\tt CLIF} are better thought to occur {\em after\/} conceptual scaling,
whereas both {\sc URC} and {\sc SOIF} specify ``raw meta-information'' 
which exists {\em before\/} conceptual scaling \cite{GaWi89}.
From the philosophical viewpoint of Concept Analysis,
conceptual scaling is an act of interpretation,
which maps raw uninterpreted data,
such as occurs in {\sc URC} or {\sc SOIF},
to a user's view.
{\sc URC} and {\sc SOIF} represent entity relations,
whereas {\tt FCIF} and {\tt CLIF} represent incidence data
between objects and attributes.
These attributes are simple structured queries
of the form {\it tag\/}\#{\it value\/},
where \# is any relational operator $+$, $\leq$, etc.
The equality operator represents {\em nominal scaling\/},
whereas the inequality operator $\leq$ represents {\em ordinal scaling\/} \cite{GaWi89}.
Through conceptual scaling, 
which often is just nominal or ordinal scaling,
we can compare {\tt FCIF} and {\tt CLIF} with {\tt URC} and {\tt SOIF}.

The {\tt OBJECT} section of the {\tt FCIF} specifies,
in a transposed fashion,
a required ordering relationship between resources:
\begin{center}
	$O_{i_1} \leq O_{i_2}$ iff $O_{i_1}$ is listed in the row indexed by $O_{i_2}$
\end{center}
This {\tt OBJECT} section can specify
both {\em generalization-specialization\/} and {\em part-whole\/}
relationships between resources.
When parts are typed,
part-whole relationships can be embedded as 
generalization-specialization relationships
--- the whole is a special case of an object which has that part.
An important example of generalization-specialization relationships
occurs in {\sc Whois}$+$$+${\sc -URC}.
Here
the instantiation relationship
between an {\sc URN} {\tt URN} and its {\sc URL} instances 
$\{ \mbox{{\tt URL}}_1, {\tt URL}_2, \ldots , \mbox{{\tt URL}}_u \}$
is a generalization-specialization relationship.
This instantiation relationship can be specified by adding a row indexed by the {\sc URN}
to the {\tt OBJECT} section of the {\tt FCIF}:
\begin{center}\begin{tabular}{l}
{\tt OBJECT} \\
\mbox{\hspace{12mm}} \verb| ... | \\
\mbox{\hspace{6mm}} \verb|URN { |${\tt URL}_1$\ ${\tt URL}_2$\verb| ... |${\tt URL}_u$\verb| }| \\
\mbox{\hspace{12mm}} \verb| ... | \\
\end{tabular}\end{center}
An important example of whole-part relationships
occurs in {\sc Harvest} {\tt SOIF}.
Here
the {\tt embedded} relationship that occurs 
between an archived directory structure and one of its summarized files
is a whole-part relationship.
This directory-file whole-part relationship can be specified 
in an analogous way
in the {\tt OBJECT} section of the {\tt FCIF}.

\subsection{Examples}

\begin{itemize}
	\item The interchange formats {\tt FCIF} and {\tt CLIF}
		subsume the Uniform Resource Characteristics ({\sc URC}s)
		of the {\sc IETF} working group on Uniform Resource Identifiers {\sc URI}s.
		The structural information in these interchange formats
		subsume Mitra's precedence rules,
		which were used by Michael Mealling \cite{Me94b}.
		As Mealling points out,
		precedence rules allow for the creation of simple {\sc URC}s
		that can be easily parsed and created by novice users.

		The example in the left side of Table~\ref{URC} of a {\sc URC}
		is taken from the Internet Draft document \cite{Me94b} of the IETF.
		This {\sc URC} contains two instances of the resource 
		whose title and author are given below the {\sc URN} attribute. 
		It describes the resource 
		entitled ``Intro to FTP and Telnet''
		by author Adam Arrowood,
		and is available 
		via anonymous ftp in postscript form 
		and 
		via http as an HTML document.
		After suitable analysis and interpretation via conceptual scaling 
		of this bibliographic record data,
		a conceptual structure such as in the right side of Table~\ref{URC}
		might be visualized.
		Here the lattice order
		from the {\sc URL} object concepts to the {\sc URN} object concept
		represents the precedence order in the {\sc URC}.
		The conceptual class labeled by the author attribute
		is distinct from the {\sc URN} object concept,
		since it represents all the resources
		which Adam Arrowood has created or authored.
		As is appropriate,
		the location is an attribute for the {\sc URL}s,
		but not the {\sc URN}.
		Table~\ref{FCIF:urc} demonstrates how 
		the {\sc URC} data on the left side in Table~\ref{URC}
		might be represented in {\tt FCIF}.
		
		From the viewpoint of Concept Analysis,
		both resources represented by {\sc URN}s 
		and 
		instances of resources represented by {\sc URL}s
		are objects,
		whereas
		meta-information in the form of
		$\pair{\mbox{multi-valued attribute}}{\mbox{value}}$
		pairs
		are (single-valued) attributes.
		Using precedence rules
		{\sc URC}s represent two kinds of relationships:
		the ``has'' or ``having'' relationship
		between a {\sc URI} and its meta-information:
		and
		the ``instantiation'' relationship
		between a resource represented by an {\sc URN}
		and one of its instances represented by a {\sc URL}.
	\item The interchange formats {\tt FCIF} and {\tt CLIF}
		subsume the Summary Object Interchange Format ({\tt SOIF})
		stream protocol of the {\sc Harvest} system.
		{\tt SOIF} specifies the interface between the {\sc Harvest} Gatherer and Broker components.
		{\tt SOIF} specifies single object summaries.
		{\tt FCIF} specifies a typed collection of {\tt SOIF}.

		Since {\tt SOIF} specifies only single objects,
		it is a special case of the {\tt FCIF} format,
		and is not able to specify order information
		either between objects or between attributes.
		Table~\ref{SOIF} demonstrates this fact.
		On the left side of Table~\ref{SOIF} is {\tt SOIF} of some type {\tt TYPE}.
		On the right side of Table~\ref{SOIF} is the corresponding {\tt FCIF} format
		for this {\tt SOIF}.
		The attributes here are the ones used to query the broker(s).
		The objects here are object summaries in the returned query results.
		We see that the {\tt SOIF} specifies mainly the incidence matrix.
	\item At the December 1994 meeting of the IETF-URI working group in San Jose,
		Stuart Weibel reported on
		``Existing Library Standards and 
		the Evolution of Uniform Resource Characteristics'',
		where he suggested
		the use of Text Encoding Initiative ({\sc TEI}) headers
		as a possible candidate for the specification of {\sc URC}s.
		The Text Encoding Initiative is a multilingual, international project
		which has developed guidelines
		for the preparation and interchange
		of electronic texts for scholarly research.
		Each {\sc TEI}-conformant text is prefixed by a header,
		which documents the electronic text.
		The use of the {\sc TEI}-header 
		as a basis for electronic text cataloging
		would allow
		digital libraries and traditional libraries to interoperate
		by integrating
		electronic resource discovery with existing 
		paper resource discovery and retrieval. 

		In the follow-up discussion of the IETF-URI working group
		Dirk Herr-Hoyman created the simple example in Table~\ref{SGML}
		for the specification of a {\sc URC}
		using {\sc TEI}-like syntax.
		The nesting here
		verifies the simple fact that
		{\sc URL}s are objects with their own attributes.
		Table~\ref{FCIF:sgml} demonstrates how the the {\sc URC} data
		in Table~\ref{SGML} might be represented in {\tt FCIF}.
	\item Each resource is of a particular type.
		That type may be standardized by the conventions of some group.
		Or,
		that type might be dependent upon 
		the individual user's purpose and current viewpoint.
		The individual user may wish to customize the resource meta-information
		in order to construct their own personal information space.

		Stuart Weibel,
		in the charter for the {\sl OCLC-NCSA\/} metadata workshop in March 1995,
		has discussed the desirability of a taxonomy of resource types.
		The complexity of resource meta-information depends upon its intended use.
		Too little information will fail to meet some purposes; 
		too much information is a burden for systems 
		and is costly to generate and maintain.
		This suggests the desirability, or even the necessity,
		for the development of a hierarchy of resource meta-information types.
		As Weibel suggests,
		resource meta-information might be promoted 
		from a simple to a more complex level 
		as the result of user demand or attention from a cataloging or archival agency.
		Levels of such a hierarchy might be defined by several criteria:
		purpose, cost, origin, etc.

		BibTeX \cite{Le79} provides a very simple example 
		for a taxonomy of types of resource meta-information.
		Different types of publications require different information:
		journal article meta-information has a volume,
		but book meta-information does not.
		For each resource type in BibTeX,
		attribute tags are divided into {\em required\/}, {\em optional\/}, and {\em ignored\/}.
		Table~\ref{bibtex:cxt} represents the BibTeX formal context,
		whose objects are the BibTeX entry types,
		and whose attributes are the \underline{required} BibTeX fields 
		(required BibTeX attribute tags).
		Figure~\ref{bibtex:lat} displays the line diagram of the lattice of classes
		for this BibTeX formal context.
		
		Here,
		the ``misc'' BibTeX entry type labels the top lattice node,
		since it is the most general type in terms of required tags (fields)
		--- it has none.
		The BibTeX ``article'' entry type 
		is more specialized than
		the BibTeX ``proceedings'' entry type,
		since in addition to ``year'' and ``title'' tags,
		it also requires ``journal'' and ``author'' tags.
		All absolutely non-required tags
		(tags not required by any type)
		label the bottom lattice node.
		This BibTeX concept lattice line diagram
		reveals an important idea:
		the discovery by conceptual analysis of new resource types
		--- the unlabelled lattice node in the center of the diagram
		represents an interesting but unspecified type,
		whose required tags are ``author'', ``title'', and ``year''.
\end{itemize}

\small
\begin{table}[htb]
\begin{center}
\begin{tabular}[t]{c}
\begin{tabular}[t]{cc}
{\footnotesize{\begin{tabular}[t]{|c|}\hline
	{\bf Formal Context Interchange Format} \\ \hline\hline \\
\begin{tabular}{l}
{\tt TYPE} \\
\mbox{\hspace{6mm}} $T$ \\
{\tt OBJECT} \\
\mbox{\hspace{6mm}} $O_1$ \verb|{| $O_{1,1}\; O_{1,2}\; \cdots\; O_{1,o_1}$ \verb|}| \\
\mbox{\hspace{6mm}} $O_2$ \verb|{| $O_{2,1}\; O_{2,2}\; \cdots\; O_{2,o_2}$ \verb|}| \\
\mbox{\hspace{6mm}} $\cdots$ \\
\mbox{\hspace{6mm}} $O_n$ \verb|{| $O_{n,1}\; O_{n,2}\; \cdots\; O_{n,o_n}$ \verb|}| \\
{\tt ATTRIBUTE} \\
\mbox{\hspace{6mm}} $A_1$ \verb|{| $A_{1,1}\; A_{1,2}\; \cdots\; A_{1,a_1}$ \verb|}| \\
\mbox{\hspace{6mm}} $A_2$ \verb|{| $A_{2,1}\; A_{2,2}\; \cdots\; A_{2,a_2}$ \verb|}| \\
\mbox{\hspace{6mm}} $\cdots$ \\
\mbox{\hspace{6mm}} $A_m$ \verb|{| $A_{m,1}\; A_{m,2}\; \cdots\; A_{m,a_m}$ \verb|}| \\
{\tt INCIDENCE} \\
\mbox{\hspace{6mm}} $O_1$ \verb|{| $A_{1,1}\; A_{1,2}\; \cdots\; A_{1,i_1}$ \verb|}| \\
\mbox{\hspace{6mm}} $O_2$ \verb|{| $A_{2,1}\; A_{2,2}\; \cdots\; A_{2,i_2}$ \verb|}| \\
\mbox{\hspace{6mm}} $\cdots$ \\
\mbox{\hspace{6mm}} $O_n$ \verb|{| $A_{n,1}\; A_{n,2}\; \cdots\; A_{n,i_n}$ \verb|}| \\
\end{tabular}
\\ \\ \hline
\end{tabular}}}
&
{\footnotesize{\begin{tabular}[t]{|c|}\hline
	{\bf Concept Lattice Interchange Format} \\ \hline\hline \\
\begin{tabular}{l}
{\tt TYPE} \\
\mbox{\hspace{6mm}} $T$ \\
{\tt GENERATOR:OBJECT} \\
\mbox{\hspace{6mm}} $C_1$ \verb|{| $O_{1,1}\; O_{1,2}\; \cdots\; O_{1,o_1}$ \verb|}| \\
\mbox{\hspace{6mm}} $C_2$ \verb|{| $O_{2,1}\; O_{2,2}\; \cdots\; O_{2,o_2}$ \verb|}| \\
\mbox{\hspace{6mm}} $\cdots$ \\
\mbox{\hspace{6mm}} $C_p$ \verb|{| $O_{p,1}\; O_{p,2}\; \cdots\; O_{p,o_p}$ \verb|}| \\
{\tt GENERATOR:ATTRIBUTE} \\
\mbox{\hspace{6mm}} $C_1$ \verb|{| $A_{1,1}\; A_{1,2}\; \cdots\; A_{1,a_1}$ \verb|}| \\
\mbox{\hspace{6mm}} $C_2$ \verb|{| $A_{2,1}\; A_{2,2}\; \cdots\; A_{2,a_2}$ \verb|}| \\
\mbox{\hspace{6mm}} $\cdots$ \\
\mbox{\hspace{6mm}} $C_p$ \verb|{| $A_{p,1}\; A_{p,2}\; \cdots\; A_{p,a_p}$ \verb|}| \\
{\tt SUCCESSOR} \\
\mbox{\hspace{6mm}} $C_1$ \verb|{| $C_{1,1}\; C_{1,2}\; \cdots\; C_{1,s_1}$ \verb|}| \\
\mbox{\hspace{6mm}} $C_2$ \verb|{| $C_{2,1}\; C_{2,2}\; \cdots\; C_{2,s_2}$ \verb|}| \\
\mbox{\hspace{6mm}} $\cdots$ \\
\mbox{\hspace{6mm}} $C_p$ \verb|{| $C_{p,1}\; C_{p,2}\; \cdots\; C_{p,s_p}$ \verb|}| \\
\end{tabular}
\\ \\ \hline
\end{tabular}}}
\end{tabular}
\\ \\
\begin{tabular}{c} \\
\begin{minipage}{10.7cm}
\begin{itemize}
	\item $O_i$ and $O_{i,o}$ are object names (strings).
	\item $A_i$ and $A_{j,a}$ are attributes {\it tag\/}\#{\it value\/}, where \# is $=$, $\leq$, etc.
	\item $C_k$ and $C_{k,s}$ are indexes (natural numbers) of conceptual classes.
	\item $x_i$ and $y_j$     are coordinates (natural numbers) of conceptual class nodes.
\end{itemize}
\end{minipage}\\
\end{tabular}
\end{tabular}
\end{center}
\caption{{\bf Interchange Formats for Faceted Resource Meta-information}}
\label{formats}
\end{table}
\normalsize


\footnotesize
\begin{table}[hbt]
\begin{center}
{\footnotesize{\begin{tabular}{|c|c|}\hline
	\rule{0pt}{8pt}{\bf Specification of a} {\sc URC}	
	& \rule{0pt}{8pt}{\bf Conceptual Structure of a} {\sc URC} \\ \hline\hline
\begin{tabular}{c}
{\scriptsize{\begin{minipage}{2.5in}
\begin{verbatim}
URN:IANA:623:oit:cs:ftp-and-telnet
Title: Intro to FTP and Telnet
Author: Adam Arrowood
URL:file://ftp.gatech.edu/pub/docs/ftp.telnet.ps
Content-Type:  text/postscript
Size: 1MB
URL:http://www.gatech.edu/oit/info/ftp.telnet.html
Content-Type: text/html
Size: 600K
Cost: US$0.25
\end{verbatim}
\end{minipage}}}
\end{tabular}
&
\setlength{\unitlength}{0.7pt}
\newcommand{\puttext}[3]{\put(#1,#2){{\mbox{\tiny$#3$\normalsize}}}}
\newcommand{\putdisk}[3]{\put(#1,#2){\circle*{#3}}}
\begin{tabular}{c}
\begin{picture}(200,210)(40,25)
	\puttext{0}{0}{{\bf }}
	\putdisk{25}{175}{7}				
	\puttext{30}{180}{\mbox{\bf size{=}large}}
	\putdisk{150}{200}{7}				
	\puttext{155}{205}{\mbox{\bf location:country{=}us}}	
	\putdisk{100}{150}{7}				
	\puttext{105}{155}{\mbox{\bf author{=}"Adam Arrowood"}}	
	\putdisk{100}{100}{7}				
	\puttext{105}{105}{\mbox{\bf title{=}"Intro to Ftp and Telnet"}}	
	\puttext{105}{95}{{\rm URN}}			
	\putdisk{50}{50}{7}				
	\puttext{55}{45}{{\rm URL1}}			
	\putdisk{150}{50}{7}				
	\puttext{155}{45}{{\rm URL2}}			
	\put(50,50){\line(2,3){100}}			
	\put(150,50){\line(0,1){150}}			
	\put(100,100){\line(-1,1){75}}			
	\put(100,100){\line(0,1){50}}			
	\put(50,50){\line(1,1){50}}			
	\put(150,50){\line(-1,1){50}}			
\end{picture}
\end{tabular}
\\ \hline
\end{tabular}}}
\end{center}
\caption{{\bf Conceptual Scaling of a} {\sc URC}}
\label{URC}
\end{table}
\normalsize


\footnotesize
\begin{table}[htb]
\begin{center}
{\footnotesize{\begin{tabular}[t]{|c|}\hline
	\rule{0pt}{8pt}{\bf Formal Context Interchange Format} \\ \hline\hline \\
\begin{tabular}{l}
{\scriptsize{\begin{minipage}{3.3in}
\begin{verbatim}
TYPE
    ????
OBJECT
    URN:IANA:623:oit:cs:ftp-and-telnet                     {
        URL:file://ftp.gatech.edu/pub/docs/ftp.telnet.ps
        URL:http://www.gatech.edu/oit/info/ftp.telnet.html
    }
    URL:file://ftp.gatech.edu/pub/docs/ftp.telnet.ps       {}
    URL:http://www.gatech.edu/oit/info/ftp.telnet.html     {}
ATTRIBUTE
    title = "Intro to FTP and Telnet" {
        author = "Adam Arrowood"
    }
    author = "Adam Arrowood"          {}
    content-type = text/postscript    {}
    content-type = text/html          {} 
    location:country = us             {}
    size = large                      {}
    file-size = 1MB                        {}
    file-size = 600K                       {}
    Cost = US$0.25                    {}
INCIDENCE
    URN:IANA:623:oit:cs:ftp-and-telnet                 {
        title = "Intro to FTP and Telnet"
        author = "Adam Arrowood"
        size = large
    }
    URL:file://ftp.gatech.edu/pub/docs/ftp.telnet.ps   {
        content-type = text/postscript
        location:country = us
    }
    URL:http://www.gatech.edu/oit/info/ftp.telnet.html {
        content-type = text/html
        location:country = us
    }
\end{verbatim}
\end{minipage}}}
\end{tabular}
\\ \\ \hline
\end{tabular}}}
\end{center}
\caption{{\bf Formal Context Interchange Format for a} {\sc URC}}
\label{FCIF:urc}
\end{table}
\normalsize


\small
\begin{table}[htb]
\begin{center}
{\footnotesize{\begin{tabular}{cc}
\begin{tabular}[t]{|c|}\hline
	\rule{0pt}{8pt}{{\bf Summary Object Interchange Format}} \\ \hline\hline \\
{\scriptsize{\begin{tabular}{l}
\verb|@ UPDATE {| \\
\verb|@| $T$ \verb|{| $URL_1\;\; A_{1,1}\; A_{1,2}\; \cdots\; A_{1,i_1}$ \verb|}| \\
\verb|@| $T$ \verb|{| $URL_2\;\; A_{2,1}\; A_{2,2}\; \cdots\; A_{2,i_2}$ \verb|}| \\
\mbox{\hspace{6mm}} $\cdots$ \\
\verb|@| $T$ \verb|{| $URL_n\;\; A_{n,1}\; A_{n,2}\; \cdots\; A_{n,i_n}$ \verb|}| \\
\verb|}| \\
\end{tabular}}}
\\ \\ \hline
\end{tabular}
&
\begin{tabular}[t]{|c|}\hline
	\rule{0pt}{8pt}{\bf Formal Context Interchange Format} \\ \hline\hline \\
{\scriptsize{\begin{tabular}{l}
{\tt TYPE} \\
\mbox{\hspace{6mm}} $T$ \\
{\tt OBJECT} \\
\mbox{\hspace{6mm}} $URL_1$ \verb|{}| \\
\mbox{\hspace{6mm}} $URL_2$ \verb|{}| \\
\mbox{\hspace{6mm}} $\cdots$ \\
\mbox{\hspace{6mm}} $URL_n$ \verb|{}| \\
{\tt ATTRIBUTE} \\
\mbox{\hspace{6mm}} $A_1$ \verb|{}| \\
\mbox{\hspace{6mm}} $A_2$ \verb|{}| \\
\mbox{\hspace{6mm}} $\cdots$ \\
\mbox{\hspace{6mm}} $A_m$ \verb|{}| \\
{\tt INCIDENCE} \\
\mbox{\hspace{6mm}} $URL_1$ \verb|{| $A_{1,1}\; A_{1,2}\; \cdots\; A_{1,i_1}$ \verb|}| \\
\mbox{\hspace{6mm}} $URL_2$ \verb|{| $A_{2,1}\; A_{2,2}\; \cdots\; A_{2,i_2}$ \verb|}| \\
\mbox{\hspace{6mm}} $\cdots$ \\
\mbox{\hspace{6mm}} $URL_n$ \verb|{| $A_{n,1}\; A_{n,2}\; \cdots\; A_{n,i_n}$ \verb|}| \\
\end{tabular}}}
\\ \\ \hline
\end{tabular}
\end{tabular}}}
\end{center}
\caption{ {\tt SOIF} {\bf and the corresponding} {\tt FCIF} }
\label{SOIF}
\end{table}
\normalsize


\footnotesize
\begin{table}[hbt]
\begin{center}
{\footnotesize{\begin{tabular}{|c|}\hline
	\rule{0pt}{8pt}{\sc TEI-SGML} \\ \hline\hline
\begin{tabular}{c}
{\scriptsize{\begin{minipage}{3.0in}
\begin{verbatim}
 
<urc>
<urn>urn:mysite.uri/myauth/11122233</urn>
<title>My really good resource</title>
<author>Ima Nutt</author>
<date>December 22, 1994</date>
<locationGroup>
<list>
<item><url>http://www.mysite.com/myresource</url>
<extent>24567 bytes</>
<format>text/html</>
</item>
<item><url>ftp://ftp.mysite.com/pub/myresource.txt</url>
<extent>12543 bytes</>
<format>text/plain</>
</item>
</list>
</locationGroup>
</urc>

\end{verbatim}
\end{minipage}}}
\end{tabular}
\\ \hline
\end{tabular}}}
\end{center}
\caption{{\bf Specification of a} {\sc URC}}
\label{SGML}
\end{table}
\normalsize

\footnotesize
\begin{table}[htb]
\begin{center}
{\footnotesize{\begin{tabular}[t]{|c|}\hline
	\rule{0pt}{8pt}{\bf Formal Context Interchange Format} \\ \hline\hline \\
\begin{tabular}{l}
{\scriptsize{\begin{minipage}{3.0in}
\begin{verbatim}
TYPE
    ????
OBJECT
    urn:mysite.uri/myauth/11122233                     {
        url:http://www.mysite.com/myresource
        url:ftp://ftp.mysite.com/pub/myresource.txt
    }
    url:http://www.mysite.com/myresource               {}
    url:ftp://ftp.mysite.com/pub/myresource.txt        {}
ATTRIBUTE
    title = "My really good resource" {
        author = "Ima Nutt"
    }
    author = "Ima Nutt"               {}
    date = "December 22, 1994"        {}
    extent = 24567bytes               {}
    format = text/html                {}
    extent = 12543bytes               {}
    format = text/plain               {}
INCIDENCE
    urn:mysite.uri/myauth/11122233                     {
        title = "My really good resource"
        author = "Ima Nutt"
        date = "December 22, 1994"
    }
    url:http://www.mysite.com/myresource               {
        extent = 24567bytes
        format = text/html
    }
    url:ftp://ftp.mysite.com/pub/myresource.txt        {
        extent = 12543bytes
        format = text/plain
    }
\end{verbatim}
\end{minipage}}}
\end{tabular}
\\ \\ \hline
\end{tabular}}}
\end{center}
\caption{{\bf Specification of a} {\sc URC}}
\label{FCIF:sgml}
\end{table}
\normalsize


\footnotesize
\begin{table}
\begin{center}
{\footnotesize{\begin{tabular}{c@{\hspace{5mm}}c@{\hspace{5mm}}c}
{\scriptsize{\begin{tabular}[t]{|r@{\hspace{2mm}}l|} \hline
	\multicolumn{2}{|c|}{{\bf objects}} \\ \hline\hline
	1 & article \\
	2 & book \\
	3 & booklet \\
	4 & inbook \\
	5 & incollection \\
	6 & inproceedings \\
	7 & manual \\
	8 & mastersthesis \\
	9 & misc \\
	10 & phdthesis \\
	11 & proceedings \\
	12 & techreport \\
	13 & unpublished \\ \hline
\end{tabular}}}
&
{\scriptsize{\begin{tabular}[t]{|r|
c@{\hspace{3pt}}c@{\hspace{3pt}}c@{\hspace{3pt}}c@{\hspace{3pt}}c@{\hspace{3pt}}
c@{\hspace{3pt}}c@{\hspace{3pt}}c@{\hspace{3pt}}c@{\hspace{3pt}}c@{\hspace{3pt}}
c@{\hspace{3pt}}c@{\hspace{3pt}}c@{\hspace{3pt}}c@{\hspace{3pt}}c@{\hspace{3pt}}
c@{\hspace{3pt}}c@{\hspace{3pt}}c@{\hspace{3pt}}c@{\hspace{3pt}}c|} \hline
	\multicolumn{21}{|c|}{{\bf incidence}} \\ \hline\hline
	           & 1 & 2 & 3 & 4 & 5 & 6 & 7 & 8 & 9 & 10 
	           & 11 & 12 & 13 & 14 & 15 & 16 & 17 & 18 & 19 & 20 \\ \hline
	1 &$\times$&$\times$&$\times$&&&&&&&&&&$\times$&&&&&&& \\
	2 &$\times$&$\times$&&&&&&&$\times$&&&&$\times$&&&&&&& \\
	3 &&$\times$&&&&&&&&&&&&&&&&&& \\
	4 &$\times$&$\times$&&&&&&&$\times$&&&&$\times$&&$\times$&&&&& \\
	5 &$\times$&$\times$&&$\times$&&&&&$\times$&&&&$\times$&&&&&&& \\
	6 &$\times$&$\times$&&$\times$&&&&&&&&&$\times$&&&&&&& \\
	7 &&$\times$&&&&&&&&&&&&&&&&&& \\
	8 &$\times$&$\times$&&&&&&&&&&&$\times$&&&&&$\times$&& \\
	9 &&&&&&&&&&&&&&&&&&&& \\
	10 &$\times$&$\times$&&&&&&&&&&&$\times$&&&&&$\times$&& \\
	11 &&$\times$&&&&&&&&&&&$\times$&&&&&&& \\
	12 &$\times$&$\times$&&&&&&&&&&&$\times$&&&&&&$\times$& \\
	13 &$\times$&$\times$&&&&&&&&&&&&&&&&&&$\times$ \\ \hline
\end{tabular}}}
&
{\scriptsize{\begin{tabular}[t]{|r@{\hspace{2mm}}l|} \hline
	\multicolumn{2}{|c|}{{\bf attributes}} \\ \hline\hline
	1 & author \\
	2 & title \\
	3 & journal \\
	4 & booktitle \\
	5 & volume \\
	6 & number \\
	7 & series \\
	8 & edition \\
	9 & publisher \\
	10 & address \\
	11 & howpublished \\
	12 & month \\
	13 & year \\
	14 & chapter \\
	15 & pages \\
	16 & organization \\
	17 & editor \\
	18 & school \\
	19 & institution \\
	20 & note \\ \hline
\end{tabular}}}
\end{tabular}}}
\end{center}
\caption{{\bf formal context for BibTeX types and required tags}}
\label{bibtex:cxt}
\end{table}
\normalsize


\begin{figure}
\begin{center}
\setlength{\unitlength}{1pt}
\newcommand{\putdisk}[3]{\put(#1,#2){\circle*{#3}}}
\newcommand{\puttext}[3]{\put(#1,#2){{\mbox{\scriptsize$#3$\normalsize}}}}
\begin{picture}(300,225)
\put(50,0){\begin{picture}(200,225)
	\puttext{0}{0}{{\bf }}
	\putdisk{100}{225}{7}					
	\puttext{105}{219}{{\rm misc}}			
	\putdisk{100}{200}{7}					
	\puttext{105}{205}{{\rm title}}			
	\puttext{105}{194}{{\rm booklet}}		
	\puttext{105}{188}{{\rm manual}}		
	\put(100,200){\line(0,1){25}}			
	\putdisk{50}{175}{7}					
	\puttext{55}{180}{{\rm year}}			
	\puttext{55}{169}{{\rm proceedings}}	
	\put(50,175){\line(2,1){50}}			
	\putdisk{150}{175}{7}					
	\puttext{155}{180}{{\rm author}}		
	\put(150,175){\line(-2,1){50}}			
	\putdisk{100}{150}{7}					
	\put(100,150){\line(-2,1){50}}			
	\put(100,150){\line(2,1){50}}			
	\putdisk{200}{150}{7}					
	\puttext{205}{155}{{\rm note}}			
	\puttext{205}{144}{{\rm unpublished}}	
	\put(200,150){\line(-2,1){50}}			
	\putdisk{0}{100}{7}						
	\puttext{-25}{105}{{\rm journal}}		
	\puttext{-25}{93}{{\rm article}}		
	\put(0,100){\line(2,1){100}}			
	\putdisk{25}{100}{7}					
	\puttext{15}{105}{{\rm institution}}	
	\puttext{15}{93}{{\rm techreport}}		
	\put(25,100){\line(3,2){75}}			
	\putdisk{50}{100}{7}					
	\puttext{55}{105}{{\rm school}}			
	\puttext{55}{93}{{\rm mastersthesis}}	
	\puttext{55}{86}{{\rm phdthesis}}		
	\put(50,100){\line(1,1){50}}			
	\putdisk{100}{100}{7}					
	\puttext{105}{105}{{\rm publisher}}		
	\puttext{105}{93}{{\rm book}}			
	\put(100,100){\line(0,1){50}}			
	\putdisk{150}{100}{7}					
	\puttext{155}{105}{{\rm booktitle}}		
	\puttext{155}{93}{{\rm inproceedings}}	
	\put(150,100){\line(-1,1){50}}			
	\putdisk{75}{75}{7}						
	\puttext{80}{81}{{\rm pages}}			
	\puttext{80}{69}{{\rm inbook}}			
	\put(75,75){\line(1,1){25}}				
	\putdisk{125}{75}{7}						
	\puttext{130}{69}{{\rm incollection}}		
	\put(125,75){\line(-1,1){25}}				
	\put(125,75){\line(1,1){25}}				
	\putdisk{100}{0}{7}						
	\puttext{105}{50}{{\rm editor}}			
	\puttext{105}{45}{{\rm organization}}	
	\puttext{105}{40}{{\rm chapter}}		
	\puttext{105}{35}{{\rm month}}			
	\puttext{105}{30}{{\rm howpublished}}	
	\puttext{105}{25}{{\rm address}}		
	\puttext{105}{20}{{\rm edition}}		
	\puttext{105}{15}{{\rm series}}			
	\puttext{105}{10}{{\rm number}}			
	\puttext{105}{5}{{\rm volume}}			
	\put(100,0){\line(-1,1){100}}			
	\put(100,0){\line(-3,4){75}}			
	\put(100,0){\line(-1,2){50}}			
	\put(100,0){\line(-1,3){25}}			
	\put(100,0){\line(1,3){25}}				
	\put(100,0){\line(2,3){100}}			
\end{picture}}
\end{picture}
\end{center}
\caption{{\bf lattice of classes for BibTeX types and required tags}}
\label{bibtex:lat}
\end{figure}
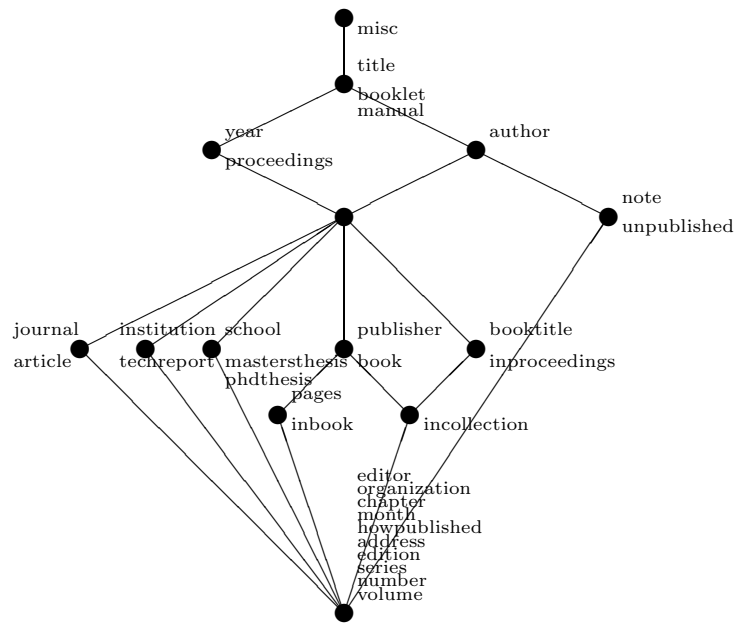


\section{Conclusions}

In this paper we have demonstrated
that Concept Analysis provides a solid foundation
for the development of a principled approach towards 
the specification of networked information resource meta-information.
We have discussed in particular,
how the meta-information in two well-known resource discovery services,
{\sc Whois}$+$$+$ and {\sc Harvest},
can both be subsumed in a more general and well-structured approach
which uses ideas from Concept Analysis.





\end{document}